\documentclass[lettersize,journal]{IEEEtran}
\usepackage{amsmath,amsfonts,amssymb}
\usepackage{algorithm}
\usepackage{algorithmicx}
\usepackage[noend]{algpseudocode}
\usepackage{array}
\usepackage{textcomp}
\usepackage{stfloats}
\usepackage{url}
\usepackage{verbatim}
\usepackage{graphicx}
\usepackage{subcaption}
\usepackage{setspace}
\usepackage{threeparttable}
\usepackage{epstopdf}
\usepackage{cite}
\usepackage{multicol}
\usepackage{stfloats}
\usepackage{color}
\usepackage{bm}
\usepackage{comment}
\begin{document}

\title{\Large{Robust Weighted Sum-Rate Maximization for Transmissive RIS Transmitter Enabled RSMA Networks}}

\author{Bojiang Li, Wen Chen, Zhendong Li, Qingqing Wu, Nan Cheng, Changle Li, Linglong Dai        
\thanks{This work is supported by National key project 2020YFB1807700, NSFC 62071296, Shanghai 22JC1404000, 20JC1416502, and PKX2021-D02.

B. Li, W. Chen, Z. Li, and Q. Wu are with the Department of Electronic Engineering, Shanghai Jiao Tong University, Shanghai 200240, China (e-mail: Li\_bojiang@sjtu.edu.cn; wenchen@sjtu.edu.cn; lizhendong@sjtu.edu.cn; qingqingwu@sjtu.edu.cn). 

N. Cheng and C. Li are with the State Key Laboratory of Integrated Services Networks, Xidian University, Xi'an 710071, China (e-mail: nancheng@xidian.edu.cn; clli@mail.xidian.edu.cn). 

L. Dai is with the Department of Electronic Engineering, Tsinghua University, Beijing 100084, China (e-mail:daill@tsinghua.edu.cn).

\textit{Corresponding author: Wen Chen}}}% <-this % stops a space

%\markboth{}%


\maketitle

\begin{abstract}
Due to the low power consumption and low cost nature of transmissive reconfigurable intelligent surface (RIS), in this paper, we propose a downlink multi-user rate-splitting multiple access (RSMA) architecture based on the transmissive RIS transmitter, where the channel state information (CSI) is only accquired partially. We investigate the weighted sum-rate maximization problem by jointly optimizing the power, RIS transmissive coefficients and common rate allocated to each user. Due to the coupling of optimization variables, the problem is non-convex, and it is difficult to directly obtain the optimal solution. Hence, a block coordinate descent (BCD) algorithm based on sample average approximation (SAA) and weighted minimum mean square error (WMMSE) is proposed to tackle it. Numerical results illustrate that the transmissive RIS transmitter with rate-splitting architecture has advantages over conventional space division multiple access (SDMA) and non-orthgonal multiple access (NOMA).  
\end{abstract}


\section{Introduction}
\IEEEPARstart{R}{ecently}, a novel multiple access method, namely, rate splitting multiple access (RSMA) has been proposed based on the space division multiple access (SDMA) and non-orthogonal multiple access (NOMA). RSMA is considered as a feasible technique which combines two decoding schemes of SDMA and NOMA, where SDMA fully treats the interference from other users as noises while NOMA fully decodes the interference \cite{9831440}. Since both two schemes can only suit for extremely weak or strong interference levels, and are sensitive to the inaccuracy of channel state information at the transmitter (CSIT), RSMA is proposed to overcome these drawback, whose excellent performance is demonstrated by the adaptive interference management strategy and the robustness under inaccurate CSIT \cite{9832611}. Specifically, RSMA enables each user's message is split into common message and private message at the transmitter side and recovered at the user side by successive interference cancellation (SIC) \cite{9771854}. Such scheme achieves a more flexible and effective interference management by partially decoding the interference and partially treating the remaining interference as noise.

On the other hand, the base stations of 5G network face the challenges of higher power consumption and deployment costs. The reconfigurable intelligent surface (RIS) is considered as an effective solution to tackle these difficulties. RIS is an array composed of a large number of low-cost passive elements; each element can be controlled to adjust the amplitude and phase shift of the incident electromagnetic wave to enable beamforming \cite{9570775}, \cite{9716123}, \cite{9887822}. The application mode of RIS can be divided into reflective RIS and transmissive RIS, where the base stations (BSs) and the users are on the same side for reflective RIS and on the different side for transmissive RIS \cite{9365009}. The communication system with reflective RIS has been well studied \cite{9983541}, \cite{9961865}. Transmissive RIS transmitter actually outperforms reflective RIS due to less feed blockage. \cite{Baixudong}. 

Based on the previous work \cite{9570775}, we introduce a transmissive RIS transmitter architecture for downlink multi-user RSMA network. The transmissive RIS transmitter can be a good alternative to multi-antenna systems and results in less power consumption and less cost in the RSMA architecture. \cite{Baixudong} Based on this system, a robust weighted sum-rate maximization problem to obtain common rate, RIS transmissive coefficients and power allocation is formulated. In order to make the problem solvable, we propose a block coordinate descent (BCD) algorithm based on sample average approximation (SAA) and weighted minimum mean square error (WMMSE) to obtain a high-quality suboptimal solution to this problem. 
\section{System Model and Problem Formulation}
\begin{figure}[ht]
\centering
\includegraphics[scale=0.2]{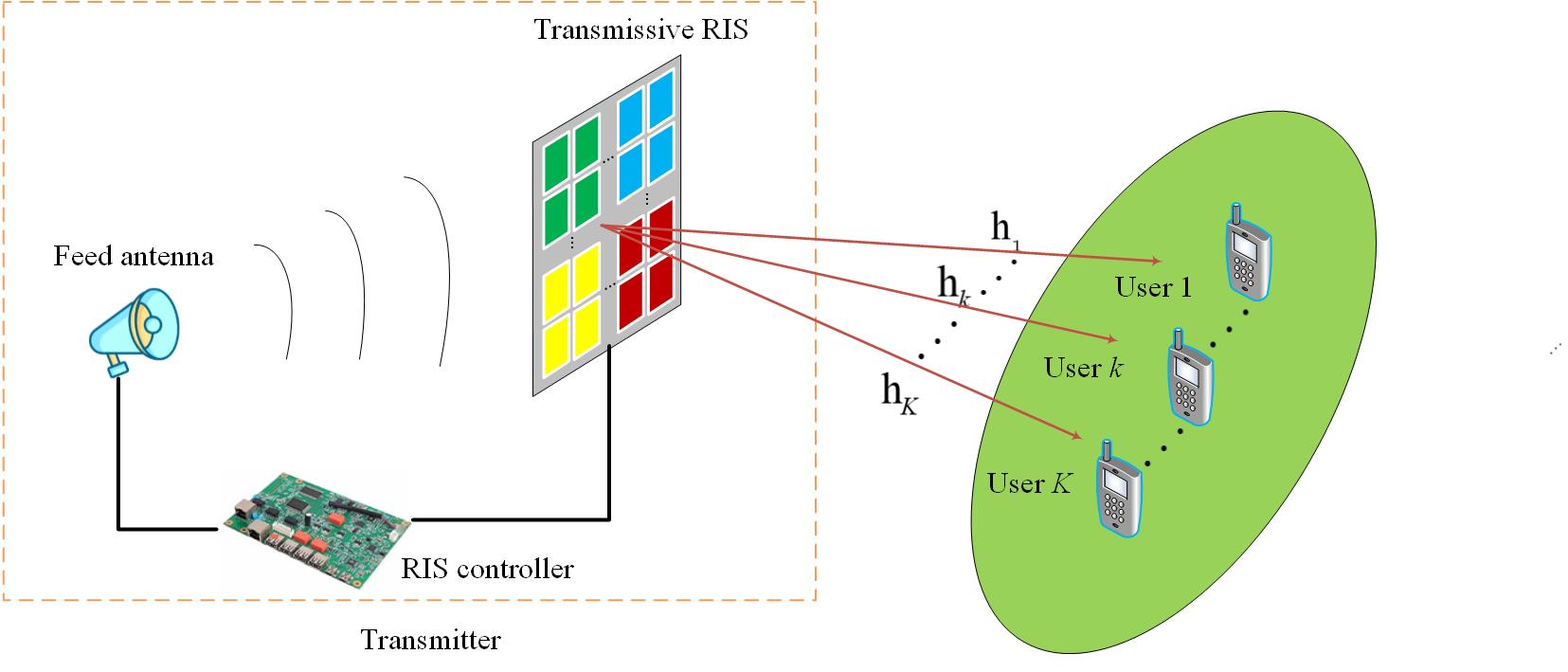}
\caption{\small{Transmissive RIS transmitter enabled RSMA architecture.}}
\label{fig_1}
\end{figure}
\subsection{System Model}
As illustrated in Fig. 1, we consider a downlink multi-user RSMA system, where the transmitter is composed of a feed antenna and a transmissive RIS panel with $N$ sub-arrays serving $K$ single-antenna users, indexed by $\mathcal{N}=\{1, \ldots, N \}$ and $\mathcal{K}=\{1, \ldots, K\}$, {\color{blue}and each sub-array includes $N_e$ elements.} Besides, the RIS controller connecting the antenna to the panel is used to control the transmissive coefficient of RIS. It is worth noting that each sub-array is equivalent to an antenna and works independently of each other controlled by the RIS controller. The signal received by the $k$-th user is defined as
\begin{equation}\label{equation1}
y_k=\mathbf{h}_k^{H} \mathbf{x}+n_k, \forall k \in \mathcal{K}, 
\end{equation}
where $\mathbf{h}_k \in \mathbb{C}^{N}$ is the channel vector bewteen the transmitter and the $k$-th user, $\mathbf{x} \in \mathbb{C}^{N}$ is the transmission signal and $n_k \sim \mathcal{C} \mathcal{N}(0,\sigma_{n, k}^2)$ is the additive white Gaussian noise (AWGN) at the $k$-th user. Without loss of generality, we assume that equal noise variances for all users, i.e., $\sigma_{n, k}^2 = \sigma_{n}^2$. It is worth noting that the channel bewteen the feed antenna and the transmissive RIS panel is not considered since the signals are transmitted from RIS to users.

In 1-layer rate-splitting architecture, each user's message $W_k$ is split into a common part $W_{c,k}$ and a private part $W_{p,k}$ at the transmitter side \cite{9831440}. The common part of each user $W_{c,k}$ can be combined into $W_{c}$. Hence, $K+1$ messages $W_c, W_{p,1} \ldots, W_{p,K}$ are sent by the transmitter instead of $K$ messages, and they are encoded into the independent data streams $\mathbf{s}=[s_c, s_1, s_2, \ldots, s_K]^{H} \in \mathbb{C}^{K+1}$, where $\mathbb{E}[\mathbf{ss}^{\mathrm{H}}]=\textbf{I}$. Let $\mathbf{p}=[p_c, p_1, \ldots, p_K]^T \in \mathbb{R}^{K+1}$ denote the transmit power vector, where $p_c$ and $p_k$ represent the power aollocated to common stream $s_c$ and the $k$-th private stream $s_k$, respectively.     

{\color{blue}In the proposed architecture, we assume that multiple elements of each sub-array can serve one stream, i.e. $N_e \geq K+1$.} Let $\mathbf{F}=[\mathbf{f}_c, \mathbf{f}_1, \ldots, \mathbf{f}_K] \in \mathbb{C}^{N \times(K+1)}$ denote the transmissive coefficients matrix of RIS, where $\mathbf{f}_c=[f_{c, 1}, \ldots, f_{c, N}]^H$ and $\mathbf{f}_k=[f_{k, 1}, \ldots, f_{k, N}]^H$ represent the RIS transmissive coefficient vector used to transmit the common stream and the $k$-th private stream, respectively. Thus, the transmission signal from the transmitter can be written as 
\begin{equation}\label{equation2}
\mathbf{x} = \mathbf{f}_c \sqrt{p_c} s_c+\sum_{j=1}^K \mathbf{f}_j \sqrt{p_j} s_j.
\end{equation} 

At the receiver side, each user decodes the common stream $s_c$ at first and extracts the corresponding common message $\hat{W}_{c,k}$ from combined $\hat{W}_c$ by treating the interference from all the private stream as noise. SIC is later used to obtain private message $\hat{W}_{p,k}$ from the corresponding private stream $s_k$ by decoding the interference stronger than the user. Then, $\hat{W}_{c,k}$ and $\hat{W}_{p,k}$ are recombined into  $\hat{W}_k$ to recover the original message of the $k$-th user. Consequently, the achievable rate of common stream $R_{c,k}$ and private stream $R_{p,k}$ for the $k$-th user can be represented as 
\begin{equation}\label{equation3}
R_{i, k}=\log _2(1+\gamma_{i, k}), \forall k \in \mathcal{K}, i \in \{c,p\}.  
\end{equation}
where $\gamma_{\mathrm{i}, k}$ is the signal-to-interference-plus-noise ratio (SINR) of the common or private stream at the $k$-th user, which can be expressed as
\begin{subequations}\label{equation4}
\begin{align}
\gamma_{c, k} &= \frac{p_c|\mathbf{h}_k^H \mathbf{f}_c|^2}{\sum\limits_{j \in \mathcal{K}} p_j|\mathbf{h}_k^H \mathbf{f}_j|^2+\sigma_n^2},   \forall k \in \mathcal{K},\\
\gamma_{p, k} & = \frac{p_k|\mathbf{h}_k^H \mathbf{f}_k|^2}{\sum\limits_{j \in \mathcal{K'}} p_j|\mathbf{h}_k^H \mathbf{f}_j|^2+\sigma_n^2},  \forall k \in \mathcal{K},
\end{align}
\end{subequations}
where $\mathcal{K'}= \mathcal{K} \backslash \{k\}$. To ensure $\textit{W}_c$ can be decoded by all users successfully, the summation of actual common rate is required not to exceed the common rate of each user, i.e., $R_c \leq R_{c, k}, \forall k \in \mathcal{K}.$ Moreover, the actual common rate of each user is $C_k$, which is a portion of $R_c$ and corresponds to the theoretical maximum common rate $R_{c,k}$. The common rate allocation $C_1, \ldots, C_K$ satisfies $\sum_{k \in \mathcal{K}}C_k = R_c$. Hence, the achievable sum rate of the $k$-th user is defined as $R_{k,sum}=C_k+R_{p, k}, \forall k \in \mathcal{K}.$

In this paper, we consider a more practical scenario where the CSI obtained by the transmitter is not accurate. For each channel vector $\mathbf{h}_k, \widehat{\mathbf{h}}_k$ denotes the estimated instantaneous channel and $\widetilde{\mathbf{h}}_k$ is the estimation error. The relationship of them can be represented as 
\begin{equation}\label{equation5}
\mathbf{h}_k = \widehat{\mathbf{h}}_k + \widetilde{\mathbf{h}}_k, \forall k \in \mathcal{K},
\end{equation}
where $\widehat{\mathbf{h}}_k$ is considered to be known in the following discussion and $\widetilde{\mathbf{h}}_k$ follows the distribution of a circularly symmetric complex Gaussian (CSCG) random vector, i.e. $\mathcal{C} \mathcal{N}(0,\sigma_{k}^2)$.

Due to the uncertainty of CSI, maximizing the instantaneous WSR may lead to transmission at undecodable rates and thus impair the system performance. Therefore, we replace WSR by the weighted ergodic sum rate (WESR) to obtain a better evaluation of the long-term WSR performance, which is defined as 
\begin{equation}\label{equation6}
\mathrm{WESR} \triangleq \sum\nolimits_{k \in \mathcal{K}} u_k\mathbb{E}_{\{\mathbf{h}_k, \widehat{\mathbf{h}}_k\}}\{R_{\mathrm{p}, k} +C_k\},
\end{equation}
where $u_k$ is the WESR weight allocated to the $k$-th user. However, due to the non-linear relationship bewteen $\mathbf{h}_k$ and $R_{i,k}$, it is difficult to get the probability density distribution of the transmission rate. We find that the long-term ergodic rate (ER) performance can be characterized by the short-term average rate (AR) performance when the number of samples is large enough. The relationship between ER and AR can be found in Eq. 8 in \cite{7555358}, where $\bar{R}_{i,k}(\widehat{\mathbf{h}}_k) \triangleq \mathbb{E}_{\mathbf{h}_k \mid \widehat{\mathbf{h}}_k}\{R_{i,k} \mid \widehat{\mathbf{h}}_k\}$ is the AR under given channel estimate $\widehat{\mathbf{h}}_k$.
\vspace{-0.76em}
\subsection{Problem Formulation}
According to the analysis above, the weighted average sum rate (WASR) maximization problem under all given channel estimates can be formulated as: 
\begin{subequations}\label{equation7}
\begin{align}
  (\text{P1}): \max_{\mathbf{p},\mathbf{F},\mathbf{c}} & \sum\limits_{k \in \mathcal{K}} u_k(\bar{R}_{p,k} + C_k),  \label{equation7a}  \\
\text{s.t.} & \sum\limits_{k \in \mathcal{K}}C_k \leq  \bar{R}_{c,k}, \forall k \in \mathcal{K},   \label{equation7b}    \\
&  C_k + \bar{R}_{p,k} \geq  R_k^{th}, \forall k \in \mathcal{K},   \label{equation7c}    \\
&  \mathbf{c} \geq  0,   \label{equation7d}  \\
&   p_c + \sum\limits_{k \in \mathcal{K}}p_k \leq  P_t, \mathbf{p} \geq  0,   \label{equation7e}  \\
&  |f_{c, n}| \leq 1,|f_{k, n}| \leq 1, \forall k \in \mathcal{K},n \in \mathcal{N},      \label{equation7f} 
\end{align}
\end{subequations}
where $\mathbf{c}=[C_1, \ldots, C_K]^T$. Constraint (\ref{equation7b}) ensures that each user is able to decode the common stream successfully. Constraint (\ref{equation7c}) guarantees that the rate of each user is no less than a certain threshold, where $R_k^{th}$ is the quality of service (QoS) threshold. Constraints (\ref{equation7d}) and (\ref{equation7e}) specify the range of the common rate allocation vector and power allocation vector, where $P_t$ represents the maximum transmit power of the RIS transmitter. Constraint (\ref{equation7f}) limits the amplitude of each RIS element. It can be seen that the problem (P1) is intractable due to the non-convexity of the objective function Eq. (\ref{equation7a}) and the constraints (\ref{equation7b}), (\ref{equation7c}). To resolve the difficulty, we apply the BCD algorithm based on SAA and WMMSE \cite{7555358}, the specific steps are detailed in next section.

\section{Solution to Optimization Problem}
To obtain AR as precise as possible, a sample average approximation method as follows is used to approach the real rates by sampling a large number of channel estimates.
\subsection{Sample Average Approximation}
Referring to the SAA method in \cite{7555358}, we can obtain the AR by averaging M rate samples, which can be defined as
\begin{subequations}\label{equation8}
\begin{align}
&\bar{R}_{i, k}^{(M)}(\widehat{\mathbf{h}}_k)  \triangleq \frac{1}{M} \sum_{m=1}^M R_{i, k}^{(m)}(\widehat{\mathbf{h}}_k), \forall k \in \mathcal{K}, i \in \{c,p\}, \\
&\bar{R}_{i, k}(\widehat{\mathbf{h}}_k) = \lim _{M \rightarrow \infty} \bar{R}^{(M)}_{i,k}(\widehat{\mathbf{h}}_k), \forall k \in \mathcal{K}, i \in \{c,p\}.
\end{align}
\end{subequations}
Consequently, the problem (P1) can be transformed into a more deterministic form to solve as follows:
\begin{subequations}\label{equation9}
\begin{align}
(\text{P2}): \max_{\mathbf{p},\mathbf{F},\mathbf{c}} & \sum\limits_{k \in \mathcal{K}} u_k(\bar{R}_{p,k}^{(M)} + C_k),     \\
\text{s.t.} & \sum\limits_{k \in \mathcal{K}}C_k \leq \bar{R}_{c,k}^{(M)}, \forall k \in \mathcal{K},     \\
 & C_k + \bar{R}_{p,k}^{(M)} \geq R_k^{th}, \forall k \in \mathcal{K},      \\
 &  (\rm{\ref{equation7d}}), (\rm{\ref{equation7e}}), (\rm{\ref{equation7f}}). 
\end{align}
\end{subequations}
It is worth noting that the coupling of optimization variables still exists in the problem (P2). Hence, we introduce a WMMSE algorithm to solve it. 
\subsection{WMMSE Algorithm}
To address problem (P2), the WMMSE algorithm is utilized to construct the rate-WMMSE relationship. For any user, the estimated common stream is first decoded and the private one is later decoded after subtracting the received common stream, which are respectively represented as $\hat{s}_{c,k} = g_{c,k}y_k$ and $\hat{s}_{p,k} = g_{p,k}(y_k- \sqrt{p_c}\mathbf{h}_k^H\mathbf{f}_{c}s_c)$, where $g_{c,k}$ and $g_{p,k}$ are the equalizer of the corresponding stream. Hence, the common and private mean square error (MSE) at the $k$-th user are respectively denoted by 
\begin{subequations}
\begin{align}\label{equation10}
\varepsilon_{c, k} &\triangleq \mathbb{E}\left\{|\hat{s}_{c, k}-s_c|^2\right\}  \notag \\
 &=|g_{c, k}|^2 T_{c, k}-2 \Re\{\sqrt{ p_c} g_{c, k} \mathbf{h}_k^{H} \mathbf{f}_c\}+1, \forall k \in \mathcal{K}, \\ 
\varepsilon_{p, k} &\triangleq \mathbb{E}\left\{|\hat{s}_{p, k}-s_k|^2\right\}  \notag \\
 &=|g_{p, k}|^2 T_{p, k}-2 \Re\{\sqrt{ p_k} g_{p, k} \mathbf{h}_k^{H} \mathbf{f}_k\}+1, \forall k \in \mathcal{K},
\end{align}
\end{subequations}
where 
\begin{subequations}\label{equation11}
\begin{align}  
T_{c, k} &=p_c\left|\mathbf{h}_k^{H} \mathbf{f}_c\right|^2+\textstyle\sum_{j \in \mathcal{K}} p_{j}\left|\mathbf{h}_k^{H} \mathbf{f}_{j}\right|^2+\sigma_k^2, \forall k \in \mathcal{K},  \\
T_{p, k} &=T_{c, k}- p_c\left|\mathbf{h}_k^{H} \mathbf{f}_c\right|^2=I_{c, k}, \forall k \in \mathcal{K},  \\
I_{p,k}  &=T_{p, k}- p_k\left|\mathbf{h}_k^{H} \mathbf{f}_k\right|^2, \forall k \in \mathcal{K}. 
\end{align}
\end{subequations}

The optimal MSE equalizers at the $k$-th user are obtained by letting $\frac{\partial \varepsilon_{p, k}}{\partial g_{p, k}}=0$ and $\frac{\partial \varepsilon_{c, k}}{\partial g_{c, k}}=0$, which are 
\begin{subequations}
\begin{align}\label{equation12}
g_{c, k}^{\mathrm{MSE}} &= \sqrt{ p_c} \mathbf{h}_k^{H} \mathbf{f}_c T_{c, k}^{-1}, \forall k \in \mathcal{K}, \\
g_{p, k}^{\mathrm{MSE}} &= \sqrt{ p_k} \mathbf{h}_k^{H} \mathbf{f}_k T_{p, k}^{-1}, \forall k \in \mathcal{K}. 
\end{align}
\end{subequations}
The minimum MSE in Eq. (10) can be rewritten by substituting Eq. (12) into it, i.e.,
\begin{equation}\label{equation13}
\varepsilon_{i, k}^{\mathrm{MSE}} \triangleq \min_{g_{i,k}}\varepsilon_{i, k} = T_{i,k}^{-1}I_{i,k}, \forall k \in \mathcal{K}, i \in \{c,p\}.
\end{equation}
It is not difficult to find $\gamma_{i,k} = (1/ \varepsilon_{i, k}^{\mathrm{MSE}})- 1,$ and $R_{i,k} =-\log_2(\varepsilon_{i, k}^{\mathrm{MSE}})$. Futhermore, the weighted mean square errors (WMSEs) are introduced to convert the non-convex rates into convex forms, which can be expressed as 
\begin{equation}\label{equation14}
\xi_{i, k}=\omega_{i, k} \varepsilon_{i, k}-\log _2\left(\omega_{i, k}\right), \forall k \in \mathcal{K}, i \in \{c,p\},
\end{equation}
where $\omega_{i, k}$ is the weight of MSE corresponding to decoding either common stream or private stream. The specific WMSE expressions for $\mathbf{p}$ and $\mathbf{F}$ are respectively expressed in Eq. (15a) and Eq. (15b).

\begin{figure*}[ht]
\begin{subequations}
\begin{align}\label{equation15}
\xi_{c, k} &= \omega_{c, k}\left|g_{c,k}\right|^2 (p_c\left|\mathbf{h}_k^{\mathrm{H}} \mathbf{f}_c\right|^2 + \sum\limits_{j=1}^Kp_j\left|\mathbf{h}_k^{H}\mathbf{f}_j\right|^2 + \sigma_k^2 )- 2\Re\{\sqrt{p_c}\omega_{c,k}g_{c,k}\mathbf{h}^{H}_k\mathbf{f}_c\}+ \omega_{c,k}- \log_2(\omega_{c,k}), \forall k \in \mathcal{K}, \\
\xi_{p, k} &= \omega_{p, k}\left|g_{p,k}\right|^2 (\sum\limits_{j=1}^Kp_j\left|\mathbf{h}_k^{H} \mathbf{f}_j\right|^2 + \sigma_k^2 )- 2\Re\{\sqrt{p_k}\omega_{p,k}g_{p,k}\mathbf{h}^{H}_k \mathbf{f}_k\}+ \omega_{p,k}- \log_2(\omega_{p,k}), \forall k \in \mathcal{K},
\end{align}
\end{subequations}
\end{figure*}

The expressions in Eq. (15) are convex with respect to $\mathbf{p}$ and $\mathbf{F}$. By optimizing WMSE weights and equalizers, the optimum WMSE can be derived by making $\frac{\partial \xi_{i,k}}{\partial g_{i,k}}=0$ and $\frac{\partial \xi_{i,k}}{\partial \omega_{i,k}}=0$, i.e.,
\begin{subequations}
\begin{align}\label{equation16}
&\omega_{i, k}^*=\omega_{i, k}^{\mathrm{MSE}} \triangleq \footnotemark[1] (\varepsilon_{i, k}^{\mathrm{MSE}})^{-1}, \forall k \in \mathcal{K}, i \in \{c,p\}, \\
&g_{i, k}^*=g_{i, k}^{\mathrm{MSE}}, \forall k \in \mathcal{K}, i \in \{c,p\}.
\end{align}
\end{subequations}
\footnotetext[1]{The constant $1/\ln{(2)}$ obtained by the derivative operation can be omitted within the error range.}Substituting this back into Eq. (14) yields the relationship as follows:
\begin{equation} \label{equation17}
\xi_{i, k}^{\mathrm{MSE}} \triangleq \min _{\omega_{i, k}, g_{i, k}} \xi_{i, k}=1-R_{i, k}, \forall k \in \mathcal{K}, i \in \{c,p\}.
\end{equation}
Thus, we establish the relationship between WMMSE and rate. The SAFs corresponding to WMSEs are shown as 
\begin{equation}\label{equation18}
\bar\xi_{i,k}^{(M)} = \frac{1}{M}\sum_{m=1}^M\xi_{i,k}^{(m)}, \forall k \in \mathcal{K}, i \in \{c,p\}, 
\end{equation}
which eliminates the effect of channel estimation errors to WMSE theoretically if $M$ tends to infinity.

For problem (P2), we reformulate the problem by replacing rates into WMMSE forms and minimizing the objective function. In detail, let $\mathbf{G}=\big\{g_{c,k}^{\mathrm{MSE}(m)},g_{p,k}^{\mathrm{MSE}(m)}\big\}$, and $\boldsymbol{\Omega} = \big\{\omega_{c,k}^{\mathrm{MSE}(m)},\omega_{p,k}^{\mathrm{MSE}(m)}\big\}$, where $m \in \mathcal{M}, k \in \mathcal{K}$. The problem is reformulated as 
\begin{subequations}
\begin{align}\label{equation19}
(\text{P3}): \min_{\mathbf{p}, \mathbf{F}, \mathbf{c}, \mathbf{G}, \mathbf{\Omega}} & \sum\limits_{k \in \mathcal{K}} u_k(\bar{\xi}_{p,k}^{(M)} - C_k)    \\
\text{s.t.} \quad & \bar{\xi}_{c,k}^{(M)} + \sum\limits_{k \in \mathcal{K}}C_k \leq 1, \forall k \in \mathcal{K},    \\
 & \bar{\xi}_{p,k}^{(M)} - C_k \leq 1 - R_k^{th}, \forall k \in \mathcal{K},     \\
 & (\rm{7d}), (\rm{7e}), (\rm{7f}). 
\end{align}
\end{subequations}
%\vspace{-2.66cm}
\subsection{Block Coordinate Descent Algorithm}
Although problem (P3) is non-convex for multi optimization variables, it is convex for a single optimization variable when the others are fixed. We utilize BCD algorithm to decouple and alternately optimize $\mathbf{p}$ and $\mathbf{F}$ by iteratively updating $\mathbf{G}$ and $\mathbf{\Omega}$ until the WASR convergences. For convenience, we simplify some parameters except $\mathbf{p}$ and $\mathbf{F}$ in Eq. (15), i.e.,
\begin{subequations}
\begin{align}\label{equation20}
t_{i, k}^{(m)} &=\omega_{i, k}^{*(m)}\left|g_{i, k}^{*(m)}\right|^2, \bar{t}_{i, k} \triangleq \frac{1}{M} \sum_{m=1}^M t_{i, k}^{(m)},   \\
\boldsymbol{\Psi}_{i, k}^{(m)} &=t_{i, k}^{(m)} \mathbf{h}_k^{(m)} \mathbf{h}_k^{H(m)}, \bar{\boldsymbol{\Psi}}_{i, k} \triangleq \frac{1}{M} \sum_{m=1}^M \boldsymbol{\Psi}_{i, k}^{(m)},   \\
\boldsymbol{\theta}_{i, k}^{(m)} &=\omega_{i, k}^{*(m)} g_{i, k}^{*(m)} \mathbf{h}_k^{H(m)}, \bar{\boldsymbol{\theta}}_{i, k} \triangleq \frac{1}{M} \sum_{m=1}^M \boldsymbol{\theta}_{i, k}^{(m)},   \\
v_{i, k}^{(m)} &=\omega_{i, k}^{*(m)}-\log _2\left(\omega_{i, k}^{*(m)}\right), \bar{v}_{i, k} \triangleq \frac{1}{M} \sum_{m=1}^M v_{i, k}^{(m)}.       
\end{align}
\end{subequations}
By applying Eq. (20) into problem (P3), a more intuitive expression upon optimizing $\mathbf{c}$, $\mathbf{p}$ and $\mathbf{F}$ can be formulated as
\begin{subequations}
\begin{align} \label{equation21}
(\text{P4}): \min_{\mathbf{p}, \mathbf{F}, \mathbf{c}, \mathbf{G}, \mathbf{\Omega}} & \sum\limits_{k \in \mathcal{K}} u_k(\Lambda_{p,k}+ \bar{t}_{p, k}\sigma_k^2- 2\mu_{p,k}    \notag \\
&+\bar{v}_{p,k} - C_k)      \\
\text{s.t.} \quad & p_c\mathbf{f}_c^{H}\bar{\boldsymbol{\Psi}}_{c, k}\mathbf{f}_c + \Lambda_{c,k}+ \bar{t}_{c, k}\sigma_k^2 - 2\mu_{c,k}    \notag  \\
&+\bar{v}_{c,k} + \sum\limits_{k \in \mathcal{K}} C_k \leq 1, \forall k \in \mathcal{K},     \\
& \Lambda_{p,k}+ \bar{t}_{p, k}\sigma_k^2 - 2\mu_{p,k}    \notag   \\
& +\bar{v}_{p,k} - C_k \leq 1- R_k^{th}, \forall k \in \mathcal{K},     \\
&  (\rm{7d}), (\rm{7e}), (\rm{7f}),
\end{align}
\end{subequations}
where $\Lambda_{i,k}=\sum_{k' \in \mathcal{K}}p_{k'}\mathbf{f}_{k'}^{H} \bar{\boldsymbol{\Psi}}_{i, k} \mathbf{f}_{k'}$, $\mu_{c,k}=\sqrt{p_c}\Re\{\bar{\boldsymbol{\theta}}_{c,k}\mathbf{f}_{c}\}$ and $\mu_{p,k}= \sqrt{p_k}\Re\{\bar{\boldsymbol{\theta}}_{p,k}\mathbf{f}_k\}$. 

The problem (P4) is convex for $\mathbf{p}$ and $\mathbf{F}$ and linear for $\mathbf{c}$, which can be alternately solved with BCD algorithm. The details of BCD algorithm are summarized in \textbf{Algorithm 1}.

\begin{algorithm}\label{algorithm}
\caption{SAA and WMMSE based BCD algorithm}\label{alg:cap}
\begin{algorithmic}[1]
\Require the convergence threshold $\epsilon$, the QoS threshold $R_k^{th}$ and the power constraint $P_t$.
\Ensure the optimal power allocation $\mathbf{p}^*$, common rate allocation $\mathbf{c}^*$ and RIS transmissive coefficients $\mathbf{F}^*$
\State Initialize ($\mathbf{p}^{[0]}, \mathbf{F}^{[0]}, \mathbf{c}^{[0]}$) and calculate $\mathrm{WASR}^{[0]}$, $n \leftarrow 0$.
\Repeat:
\State $\mathbf{G}^{[n]}$ and $\boldsymbol{\Omega}^{[n]}$ can be obtained by Eq. (16).
\State Update $\bar{t}_{i, k}^{[n]}$, $\bar{\boldsymbol{\Psi}}_{i, k}^{[n]}$, $\bar{\boldsymbol{\theta}}_{i, k}^{[n]}$, $\bar{v}_{i, k}^{[n]}$ based on Eq. (20).
\State Solve the problem (P4) alternately based on the parameters obtained above, specifically: 
\State P4.1: $\mathbf{p}^{[n+1]}$ is obtained by fixing $\mathbf{c}^{[n]}$, $\mathbf{F}^{[n]}$.
\State P4.2: $\mathbf{F}^{[n+1]}$ is obtained by fixing $\mathbf{c}^{[n]}$, $\mathbf{p}^{[n+1]}$.
\State P4.3: $\mathbf{c}^{[n+1]}$ is obtained by fixing $\mathbf{p}^{[n+1]}$, $\mathbf{F}^{[n+1]}$.
\State Update iteration $n \leftarrow n+1$.
\Until $|\mathrm{WASR}^{[n+1]} - \mathrm{WASR}^{[n]}|< \epsilon$   \\
\Return $\mathbf{p}^*=\mathbf{p}^{[n+1]}$, $\mathbf{F}^*=\mathbf{F}^{[n+1]}$, $\mathbf{c}^*=\mathbf{c}^{[n+1]}$
\end{algorithmic}
\end{algorithm}

In each iteration of \textbf{Algorithm 1}, it takes two steps to solve the problem: \textit{1)} the original non-convex problem (P1) is transformed into the convex problem (P4) by SAA and WMMSE operations, which has the complexity of $\mathcal{O}(K^2M)$. \textit{2)} the convex problem (P4) is solved by the interior point method, which has the complexity of $\mathcal{O}(K^{3.5})$. Hence, the complexity of \textbf{Algorithm 1} in each iteration is $\mathcal{O}\Big(\text{max}(K^2M,K^{3.5})\Big)$. Since $M$ is determined and considered as a fixed value, the total complexity of \textbf{Algorithm 1} is $\mathcal{O}\Big(\log(\epsilon^{-1})K^{3.5}\Big)$.

Suppose ${\mathbf{p}^{(r)}},{\mathbf{F}^{(r)}},{\mathbf{c}^{(r)}}$ as the r-th iteration solution of the \textbf{Algorithm 1}. The objective function is denoted by $\mathcal{P}\left(\mathbf{p}^{(r)},\mathbf{F}^{(r)},\mathbf{c}^{(r)}\right)$. Follow the step 6,7 and 8 of the algorithm, we can get
\begin{subequations}
\begin{align}\label{equation22}
&\mathcal{P}\left(\mathbf{p}^{(r)},\mathbf{F}^{(r)},\mathbf{c}^{(r)}\right) \geq \mathcal{P}\left(\mathbf{p}^{(r+1)},\mathbf{F}^{(r)},\mathbf{c}^{(r)}\right).   \\
& \mathcal{P}\left(\mathbf{p}^{(r+1)},\mathbf{F}^{(r)},\mathbf{c}^{(r)}\right) \geq \mathcal{P}\left(\mathbf{p}^{(r+1)},\mathbf{F}^{(r+1)},\mathbf{c}^{(r)}\right).     \\
& \mathcal{P}\left(\mathbf{p}^{(r+1)},\mathbf{F}^{(r+1)},\mathbf{c}^{(r)}\right) \geq \mathcal{P}\left(\mathbf{p}^{(r+1)},\mathbf{F}^{(r+1)},\mathbf{c}^{(r+1)}\right).
\end{align}
\end{subequations}
Based on above, we can get
\begin{equation}\label{equation23}
\mathcal{P}\left(\mathbf{p}^{(r)},\mathbf{F}^{(r)},\mathbf{c}^{(r)}\right) \geq \mathcal{P}\left(\mathbf{p}^{(r+1)},\mathbf{F}^{(r+1)},\mathbf{c}^{(r+1)}\right),
\end{equation}
which ensure the convergence of \textbf{Algorithm 1} due to the fact that the objective function has a finite lower bound.
\section{Numercial Results}

\begin{figure*}[ht]
\centering
\begin{minipage}[ht]{0.25\linewidth}
\includegraphics[width=4.8cm]{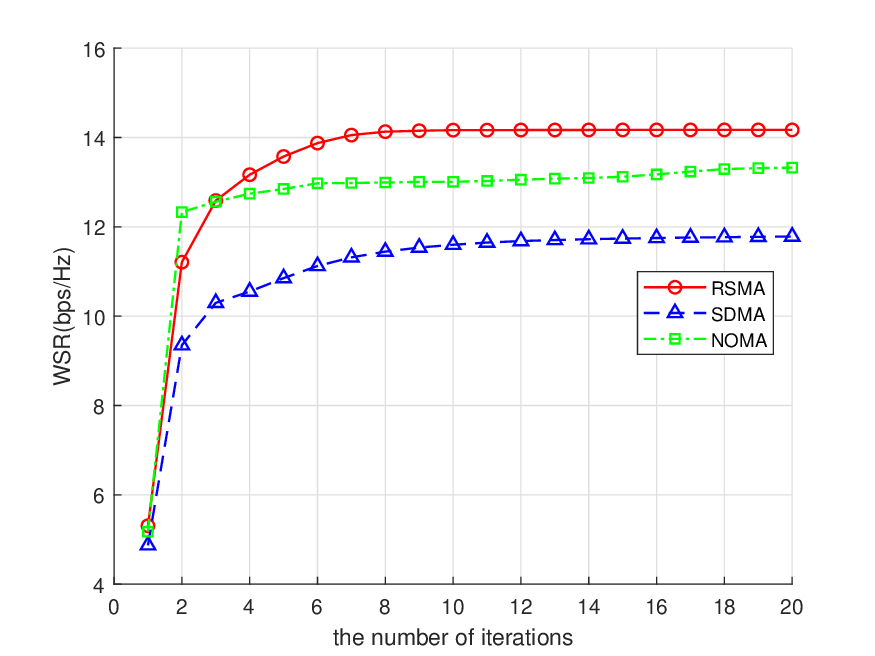}
\caption{\small{The convergence of the proposed RSMA-based algorithm compared with SDMA and NOMA}}
\label{fig_2}
\end{minipage}
\begin{minipage}[ht]{0.24\linewidth}
\centering
\includegraphics[width=4.8cm]{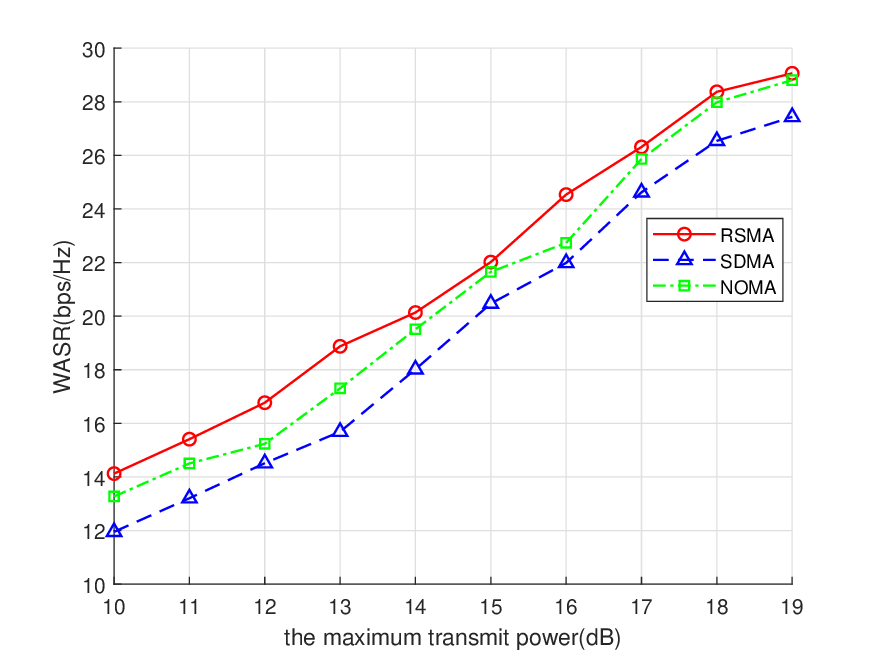}
\caption{\small{The realtionship bewteen the maximum transmit power and WASR in three architectures}}
\label{fig_3}
\end{minipage}
\begin{minipage}[ht]{0.24\linewidth}
\centering
\includegraphics[width=4.8cm]{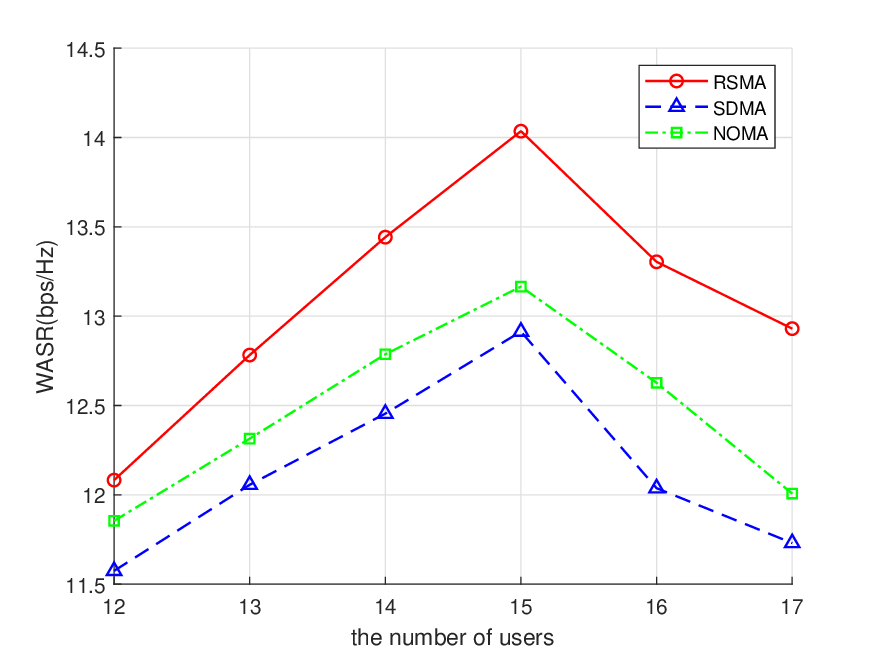}
\caption{\small{The realtionship bewteen the number of users and WASR in three architectures}}
\label{fig_4}
\end{minipage}
\begin{minipage}[ht]{0.25\linewidth}
\centering
\includegraphics[width=4.8cm]{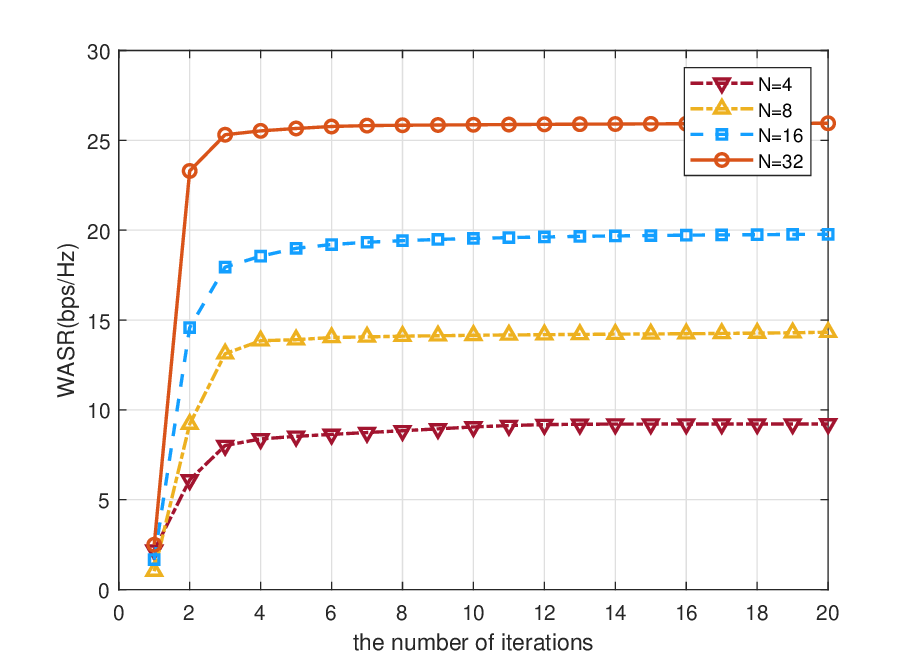}
\caption{\small{The WASR of the proposed RSMA-based algorithms for different number of RIS sub-arrays}}
\label{fig_5}
\end{minipage}
\end{figure*}
  In this section, we validate the effectiveness of the proposed RSMA-based optimization algorithm by simulations with CVX. In the simulations, $\sigma^2_k=1$. The number of channel samples is set to $M=500$. {\color{blue}Each sub-array includes $N_e=32$ elements and the total elements are 256. The channel follows a Rayleigh distribution, where the signal attenuation at a reference distance of 1 m and the pathloss exponents are set as 30 dB and 3. The
distance bewteen the transmissive RIS transmitter and users follows a uniform distribution from 1 to 100m.} The QoS threshold $R_k^{th}$ is set to 0.1 bps/Hz for each user. Each user has the same priority in communication, which means $u_k=1$ for all users. 

{\color{blue}In Fig. \ref{fig_2}, the number of RIS sub-arrays is set to $N=8$ and the number of users is set to $K=15$.} We first evaluate the convergence of the proposed RSMA-based algorithm compared with conventional SDMA and NOMA architecture. The initialization of each user's power in RSMA and SDMA architectures follows the uniform distribution at the maxium transmit power $P_t=10$dB, while in NOMA it depends on the channel strength of each user \cite{9531372}. That's why the performance of NOMA is the best in the early iterations. As the iteration continues, the performance of NOMA suffers due to the uncertain channel and performs worse than RSMA when coming to the convergence. It demonstrates that our proposed RSMA-based algorithm performs better than NOMA and SDMA when the channel acquisition is inaccurate.

{\color{blue}In Fig. \ref{fig_3}, we set $N=8$ and $K=15$.} As a complement, we analyze how the maxium transmit power, which is the most important factor constrainting WASR, affects the WASR. Although the WASR grows approximately linearly with the maximum transmit power as we expected, it may come to the peak constrainted by the real scenarios.

{\color{blue}In Fig. \ref{fig_4}, we set $N=8$ and $P_t=10$dB.} We study the realtionship bewteen the number of users and WASR in three architectures. At the initial stage, the actual transmit power is lower than the maximum transmit power, so the WASR grows like Fig. \ref{fig_3}. When the transmit power reaches the peak, the WASR won't continue to increase but decrease due to the increasing interference levels. It proves that our proposed algorithm has better interference immunity compared to others.

{\color{blue}In Fig. \ref{fig_5}, we set $K=15$ and $P_t=10$dB.} Fig. \ref{fig_5} compares the WASR under different number of RIS sub-arrays. It illustrates that the increase of RIS sub-arrays significantly improves the WASR. As mentioned above, the transmissive RIS transmitter is equivalent to a multi-antenna system, where the system capacity and the communication performance can be improved with the increase of sub-arrays. 

\section{Conclusion}
In this paper, we propose a rate-splitting multi-user access architecture with the transmissive RIS transmitter where the CSI bewteen the users and the transmitter is estimated inaccurately. A weighted sum-rate maximization problem on optimizing power allocation, common rate allocation and RIS transmissive coefficients based on RSMA architecture is mainly investivated. To deal with the non-convexity of the problem caused by the coupling of optimization variables, the SAA and WMMSE methods are utlized to transformed into a convex problem and a BCD algorithm is used for the solution to the problem. Numerical results validate that our proposed algorithm performs better compared with SDMA and NOMA architecture, and the transmissive RIS transmitter has more advantages than the conventional multi-antenna system.

\bibliographystyle{ieeetr}
\bibliography{RSMA-RIS}

\end{document}